\documentclass[twocolumn,aps,prl,superscriptaddress,longbibliography]{revtex4-2}

\usepackage{graphicx,graphics,epsfig}
\usepackage{soul,xcolor}
\usepackage{dcolumn}
\usepackage{bm}
\usepackage{amsmath,amsfonts,amssymb}
\usepackage{subfigure}                
\usepackage{times}                     
\usepackage{bbm}
\usepackage[normalem]{ulem}
\usepackage{hyperref}
\usepackage{mathtools}
\usepackage{braket}
\usepackage{comment}
\usepackage{tabularray}
\usepackage{orcidlink}

\usepackage[colorinlistoftodos,prependcaption]{todonotes}
\newcommand{\ie}{{\it i.e.,} }

\setstcolor{red}

\begin{document}

\title{Non-Exponential Decay in Finite Photonic Waveguide Arrays}

\author{Florian H. Huber\,\orcidlink{0000-0002-2928-1032}}
    \email{florian.huber@tu-berlin.de}
 	\affiliation{\LMU}
    \affiliation{\MPQ}
    \affiliation{\MCQST}
    \affiliation{\TUB}
    
\author{Benedikt Braumandl\,\orcidlink{0009-0008-4787-3526}}
    \affiliation{\LMU}
    \affiliation{\MPQ}
    \affiliation{\MCQST}
    \affiliation{\TUM}

\author{Johannes Kn{\"o}rzer\,\orcidlink{0000-0002-7318-3018}}  
    \thanks{Current Address: Department of Physics, ETH Zürich, CH-8093 Zurich, Switzerland}
 	\affiliation{\ETHZ}

\author{Jonas Himmel\,\orcidlink{0009-0007-2943-5197}}
 	\affiliation{\UR}

\author{Carlotta Versmold\,\orcidlink{0000-0001-7958-5707}}
 	\affiliation{\LMU}
    \affiliation{\MPQ}
    \affiliation{\MCQST}

\author{Robert H.~Jonsson\,\orcidlink{0000-0003-0295-250X}}
 	\affiliation{\nordita}

\author{Alexander Szameit\,\orcidlink{0000-0003-0071-6941}}
 	\affiliation{\UR}

\author{Jasmin Meinecke\,\orcidlink{0000-0003-0616-6653}}
    \email{jasmin.meinecke@tu-berlin.de}
 	\affiliation{\LMU}
    \affiliation{\MPQ}
    \affiliation{\MCQST}
    \affiliation{\TUB}

\newcommand{\ETHZ}{Institute for Theoretical Studies, ETH Zurich, Scheuchzerstrasse 70, 8006 Zurich, Switzerland}

\newcommand{\LMU}{Fakult{\"a}t f{\"u}r Physik, Ludwig-Maximilians-Universit{\"a}t M{\"u}nchen, Schellingstr. 4, 80799 M{\"u}nchen, Germany}

\newcommand{\MPQ}{Max-Planck Institut f{\"u}r Quantenoptik, Hans-Kopfermann-Str. 1, 85748 Garching, Germany}

\newcommand{\MCQST}{Munich Center for Quantum Science and Technology (MCQST), Schellingstr. 4, 80799 M{\"u}nchen, Germany}

\newcommand{\TUB}{Institute of Solid State Physics, Technische Universit{\"a}t Berlin, Hardenbergstraße 36, 10623 Berlin, Germany}

\newcommand{\TUM}{Department of Physics, Technische Universit{\"a}t M{\"u}nchen, James-Franck-Str. 1, 85748 Garching, Germany}

\newcommand{\nordita}{Nordita, Stockholm University and KTH Royal Institute of Technology, Hannes Alfv\'ens v\"ag 12, SE-106 91 Stockholm, Sweden}

\newcommand{\UR}{Institut für Physik, Universität Rostock, Albert-Einstein-Straße 23, 18059 Rostock, Germany}

\begin{abstract}
Open quantum-system dynamics can follow exponential decay, non-exponential relaxation, or oscillatory dynamics, depending on the system–environment coupling.
We study a lattice with a boundary defect that transitions between these regimes, controlled by a single parameter.
Extending the exact solution to the oscillatory case, we establish a unified theory confirmed by experiments in integrated waveguide arrays.
We characterize finite-size effects by comparing analytics, numerics, and data.
This provides a benchmark for emulating infinite systems and studying open systems in photonic lattices.

\end{abstract}

\maketitle

\begin{figure}[b]
    \centering
    \includegraphics[width=0.85\columnwidth]{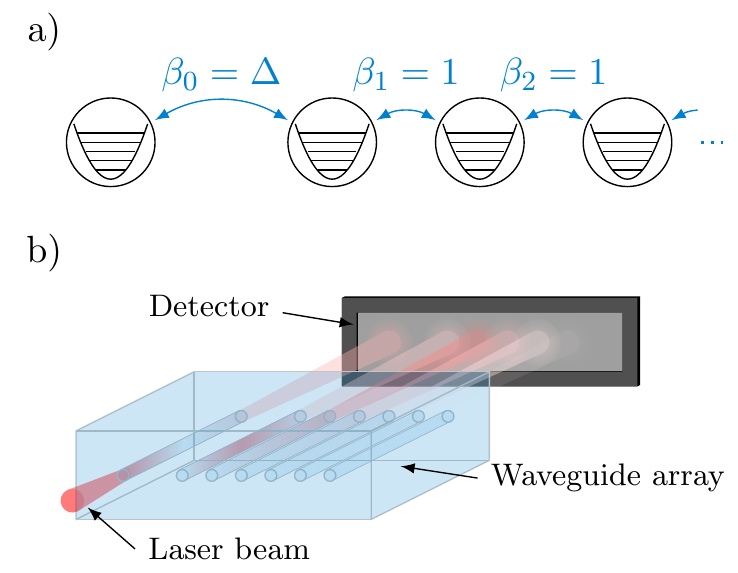}
    \caption{ Illustration a) of the underlying tight-binding model with hopping amplitudes $\beta_i$, and b) of the experimental platform used to examine and implement the dynamics of the theoretical model.
    }
    \label{fig:TB_WGA}
\end{figure}

Tight-binding models with boundary defects provide a minimal setting to study decay in open quantum systems.
A single defect at the edge of a lattice gives rise to a variety of dynamical behaviors.
These include exponential decay, non-exponential relaxation, and persistent oscillations, all governed by the coupling strength between defect and lattice boundary~\cite{longhi2006nonexponential}.
While exponential decay is the generic feature of unstable quantum states~\cite{weisskopf1930berechnung}, deviations arise in structured environments or under coherent feedback~\cite{seke1988deviations,taj2008, peshkin2014, garmon2019, kumlin2020, guoOscillatingBoundStates2020,gonzales-tuleda2017}.

To explore these effects experimentally, photonic waveguide arrays (WGAs) offer a natural realization of tight-binding models, where coupling strengths and boundary conditions can be precisely engineered.
WGAs have become a versatile platform for simulating quantum lattice systems in controlled, scalable settings~\cite{christodoulides2003discretizing,chenTightbindingModelOptical2021,joglekar2013optical,grafe2020integrated}.
In these systems, light propagation along the longitudinal waveguide axis effectively maps to time evolution under a tight-binding Hamiltonian, enabling direct access to site-resolved dynamics~\cite{aspuru-guzik2012photonic}.
This optical realization of lattice models has been successfully exploited in a wide range of experiments, including studies of topological photonic states~\cite{blanco-redondo2016topological,wang2019direct,kang2023topological}, Bloch oscillations~\cite{morandotti1999experimental}, coherent quantum transport~\cite{biggerstaff2016enhancing} and quantum walks~\cite{peruzzo2010quantum}, localization in the presence of disorder~\cite{lahini2008anderson}, and $\mathcal{PT}$-symmetric lattices~\cite{miri2012optical}.

Since small deviations in target and actual model parameters lead to accumulated simulation errors over time, accurate simulation of site-resolved quantum dynamics places high demands on the reproducibility and precision of the fabrication of WGAs.
Besides manufacturing imperfections, boundary effects of the finite extent of WGAs introduce additional deviations from idealized theoretical models, which often assume semi-infinite or infinite systems.
Modeling an infinite system by a finite implementation may alter decay processes, modify bound-state formation or result in deviations between long-time behavior of the actual and simulated systems.
For a large class of models, the validity of this truncation can be certified by appropriate error bounds~\cite{woodsSimulatingBosonicBaths2015,woods2016dynamical,trivediConvergenceGuaranteesDiscrete2021,jonssonChainmappingMethodsRelativistic2024}.

To assess the fidelity of waveguide-based simulations, especially in regimes sensitive to small imperfections, we study the photonic realization of a solvable lattice model exhibiting distinct dynamical behaviors controlled by a single tunable parameter.
Non-exponential decay through tunneling has been theoretically described and analyzed in a minimal tight-binding model featuring a boundary defect, which admits an exact solution and reveals transitions between exponential decay, persistent oscillations and non-exponential decay dynamics depending on a single coupling parameter~\cite{longhi2006nonexponential}.

In this work, we experimentally realize this boundary-defect model using femtosecond-laser-written photonic WGAs, where the coupling profiles are informed by the target model parameters.
A schematic illustration of the model and its optical implementation is shown in Fig.~\ref{fig:TB_WGA}.
By systematically varying the coupling strength at the boundary, we access the full range of dynamical regimes predicted by the model.
In addition to implementing the known analytic solution for the decaying regime, we extend the theory to cover the previously untreated parameter range characterized by coherent oscillations at the boundary site.
This completes the analytic description of the lattice model and enables a direct comparison with experimental data across all regimes.
We benchmark our photonic implementation against these exact theoretical predictions using coherent input states.
This allows us to quantitatively assess the precision of the waveguide-based simulation and evaluate how accurately finite-size arrays can replicate the behavior of an idealized semi-infinite system.
To explain additional shortcomings of the platform we compare our findings with a numerical simulation of the light field propagation inside the waveguide system by reconstructing the refractive index profile followed by eigenmode expansion (EME) simulation which is not constrained by the weak-coupling assumption.

\textit{Implementation}.\textemdash
In our experiment, we use femtosecond-laser-written WGAs in fused silica~\cite{szameit2007, szameit2010} to implement a one-dimensional lattice with tunable coupling at the boundary.
Evanescent coupling causes light to gradually spread to neighboring sites while the strength of this coupling is controlled by the transverse spacing of the waveguides.

\begin{figure}
    \includegraphics{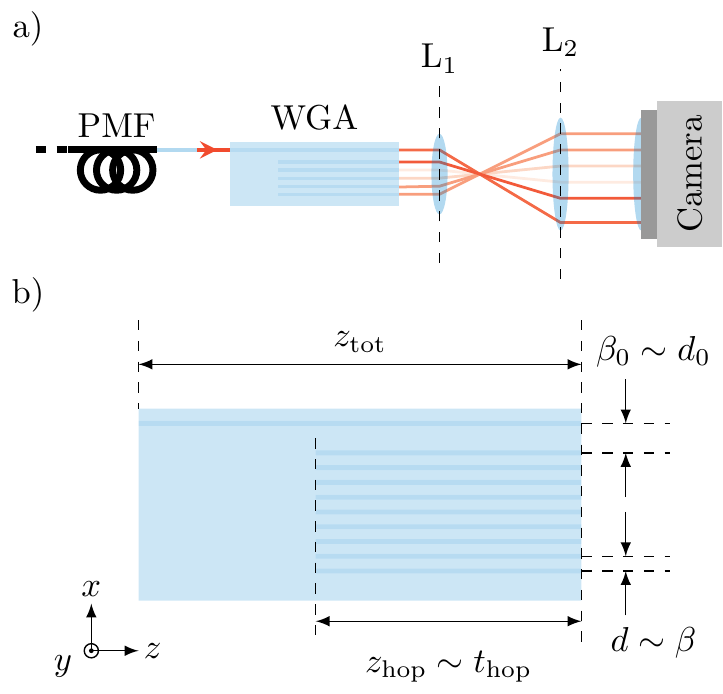}
    \caption{Sketch of the setup used in the experiment. 
    A laser beam is coupled into the waveguide array (WGA) via a polarization maintaining fiber (PMF). 
    The output facet of the array is enlarged using magnification optics (L$_1$ and L$_2$) before reaching a camera. 
    Panel b) shows an illustration of the WGA with all relevant quantities.
    The total length $z_\mathrm{tot}$ of the system is the same for all measured arrays.
    To vary the evolution time, we simply change the stretch $z_\mathrm{hop}$ over which all waveguides are present.
    The hopping amplitudes $\beta_0$ and $\beta$ are related to the separations $d_0$ and $d$ of adjacent waveguides, respectively.}
    \label{fig:setupsketch}
\end{figure}

Our experimental setup is illustrated in Fig.~\ref{fig:setupsketch}a). Each array consists of $N=10$ waveguides.
A laser beam is coupled into the first waveguide via a polarization-maintaining fiber. 
At the output facet, the intensity distribution is magnified by a factor of $55.56$ using lenses (L$_1$, L$_2$) in a 4f configuration and recorded on a camera.
For calibration, to ensure correct focusing of the magnification optics, a beam splitter is used behind L$_1$ to illuminate the facet with incoherent, white light to bring the output plane of the waveguide into focus. After calibration the beam splitter is removed to avoid photon loss and optical aberrations.

A key feature of photonic WGAs is that the longitudinal direction of light propagation corresponds to time evolution in an equivalent lattice model. 
As a result, each array of a specific length effectively encodes the system's state at a specific evolution time. 
To reconstruct the full dynamics, we fabricate a set of arrays with varying hopping distance $z_\mathrm{hop}$ while keeping the total chip length fixed, as shown in Fig.~\ref{fig:setupsketch}b). 
Each array thus acts as a snapshot of the system at a particular point in its evolution.
In our implementation, we realize an array with uniform coupling between all waveguides except at the boundary, where the first coupling is independently tunable.
We present data of three different array sets with different coupling strengths listed in Table~\ref{tab:measurementdata} and discussed in detail later.
This enables a controlled study of boundary effects and their impact on the observed dynamics.

For probing the dynamics, we use a coherent laser beam as input.
This enables high signal-to-noise detection and precise control over the initial conditions.
In linear systems, the evolution of classical coherent states follows the same equations as that of single photons, which is a general result established by Glauber~\cite{glauber1966}.
This correspondence allows us to benchmark the underlying quantum dynamics using coherent light, while retaining direct interpretability in the tight-binding framework introduced in the following section.

\textit{Model}.\textemdash
The propagation of light in the WGA is well approximated by a tight-binding model with site-dependent nearest-neighbor couplings~\cite{chenTightbindingModelOptical2021}.
This framework provides a minimal yet analytically tractable description of the system, allowing us to study features such as coherent oscillations and non-exponential decay in a controlled setting~\cite{rojas2014, szameit2007}.

We consider a one-dimensional chain of $N$ coupled sites, where the amplitude at each site corresponds to the optical field in a given waveguide.
The model includes on-site energies $\alpha_i$ and hopping amplitudes $\beta_i$ between adjacent sites.
Within the single-band approximation, it is described by the Hamiltonian
\begin{equation}
    \label{eq:ham}
    \hat{H} = \sum\limits_{i=0}^{N-1}\alpha_i\hat{a}_i^\dagger\hat{a}_i + \sum\limits_{i=0}^{N-2} \beta_i\left(\hat{a}_i^\dagger\hat{a}_{i+1}+\mathrm{h.c.}\right),
\end{equation}
where the operators $\hat{a}_i^\dagger$ and $\hat{a}_i$ create and annihilate an excitation at site $i$, and satisfy bosonic commutation relations $[\hat{a}_i,\hat{a}_j^\dagger] = \delta_{ij}$, $[\hat{a}_i,\hat{a}_j] = 0$.
In our implementation, we assume identically shaped waveguides ($\alpha_i=\alpha$).
Also, with exception of the first two sites, we assume equidistant spacing between neighboring sites ($\beta_i=\beta$ for $i>0$) and introduce $\Delta=\beta_0/\beta$ as the ratio of the two  coupling strengths. 
Note that we later use $\hat H$ to model the propagation of light along the $z$-axis of the WGA (see Fig.\ref{fig:setupsketch}). Hence, the coupling constants $\alpha,\beta$ then have units of inverse length.

To analyze the time evolution numerically, we initialize the system with a single excitation localized at the first site,
\begin{equation}
    \ket{\psi(t=0)} = \ket{1, 0, 0, \dots},
    \label{eq:init}
\end{equation}
and evolve the state under the time-evolution operator $\hat{U}(t) = \exp(-i \hat{H} t)$.
This models the injection of light into the first waveguide in the experiment. Since the system is linear and describable by a single-excitation subspace, the evolution is equivalent for both single photons and coherent states~\cite{glauber1966}, and the theoretical amplitude $c_0(t)$ for the first site can be directly compared with the measured intensity.

As shown in Ref.~\cite{longhi2006nonexponential}, the semi-infinite system admits closed-form expressions for $c_0(t)$, which reveal transitions between exponential decay, persistent oscillations, and non-exponential dynamics depending on the value of $\Delta$. In particular, the probability amplitude for finding the first site ($i=0$) in the excited state at time $t$ is given by
\begin{equation}
    \label{eq:int}
    c_0(t) = \frac{1}{2\pi i}\oint_{\sigma}dz\exp\left(it\left(z+\frac{1}{z}\right)\right)\frac{z^2-1}{z(z^2+\gamma^2)},
\end{equation}
where $\gamma := \sqrt{|1-\Delta^2|}=\sqrt{|1-(\beta/\beta_0)^2|}$,
and the contour $\sigma$ is a closed loop in the complex plane enclosing all residues of the integrand.
Using the residue theorem, one finds concise expressions for $\Delta < 1$, $\Delta = 1$ \cite{longhi2006nonexponential} and for the previously unexplored parameter regime of $\Delta > 1$:
\begin{equation}
c_0(t) = 
\begin{cases}
\frac{A}{2}\exp\left({-\Omega t}\right)+S_{<}(t), & \Delta < 1 \\
J_1(2t)/t, & \Delta = 1 \\
A \cos\left(\Omega t\right)+S_{>}(t), & \Delta > 1
\end{cases}\quad,
\label{eq:c0}
\end{equation}
where $S_<(t)$ and $S_>(t)$ modify the exponential decay and persistent oscillations of the respective cases to some extent.
Their detailed form is given in the End Matter.

In the limit $\Delta \rightarrow 0 $, the term $S_<(t) \sim \Delta^2$, \emph{i.e.}, only the exponential decay in $c_0(t)$ remains~\cite{longhi2006nonexponential}.
For $\Delta > 1$, persistent oscillatory behavior with an initial decay can be observed.
The amplitude and frequency of these oscillations are given by $A \coloneqq (\Delta^2-2)/(\Delta^2-1)$ and $\Omega \coloneqq \Delta^2/\sqrt{|\Delta^2-1|}$.

\textit{Finite-size effects}.\textemdash
We emphasize that the analytic expressions discussed above were derived for a semi-infinite chain. In practice, both numerical simulations and experimental implementations are necessarily confined to finite-sized arrays. To assess how accurately such truncated systems reproduce the ideal dynamics, we compare the evolution of finite arrays to that of a much larger reference system. This allows us to quantify finite-size effects and determine the validity range of the semi-infinite approximation in our setting.

\begin{figure*}
    \centering
    \includegraphics[width=0.75\textwidth]{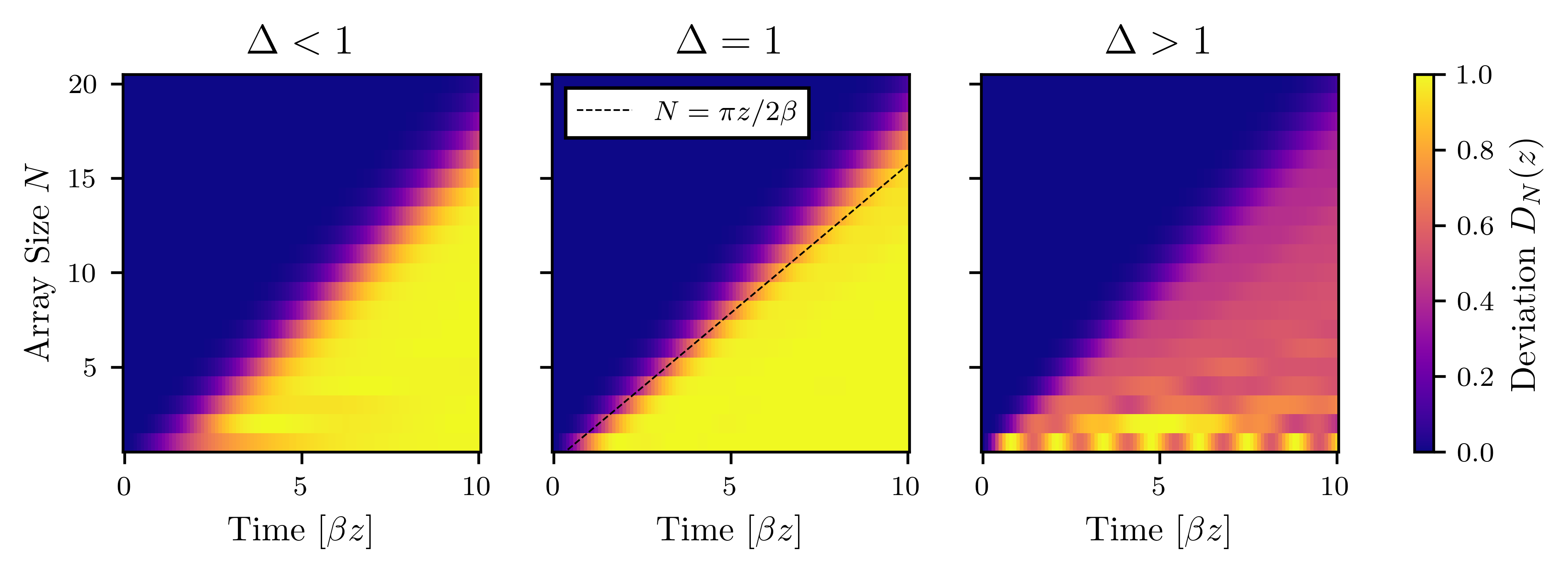}
    \caption{Illustration of deviations from the dynamics of a semi-infinite lattice due to the finite size of the WGA. Coupled mode simulations of the evolution of large arrays ($N > 500$) is compared with that of truncated arrays containing only $N$ waveguides. For small $N$, noticeable deviations $D_N(z)$ from the ideal behavior appear at earlier times with subtle yet distinct differences across the different coupling regimes. In the uniform coupling case ($\Delta = 1$), we additionally compare the numerical results to the characteristic inversion length for a pair of coupled waveguides, $z_s \sim \pi/2\beta$ represented by the black dashed line \cite{pätzold2017}, denoting the distance after which full energy transfer from one site to a neighboring site occurs. After $N$ times the most outer waveguide is reached.}
    \label{fig:deviation}
\end{figure*}

To this end, we define the deviation
\begin{equation}
    D_N(t) = 1 - \left| \braket{ \psi_N^\mathrm{trunc}(t) | \psi^\mathrm{full}(t) } \right|^2,
    \label{eq:overlap}
\end{equation}
which measures the squared overlap between the state $\ket{\psi_N^\mathrm{trunc}(t)}$ of a truncated array with $N$ sites and the state $\ket{\psi^\mathrm{full}(t)}$ of a large reference array with $N_{\mathrm{full}}$ sites, where $N_{\mathrm{full}} \gg N$.
In this comparison, the truncated state is zero-padded to match the dimension of the full system.
As shown in Fig.~\ref{fig:deviation}, the deviation $D_N(z)$ remains small across all considered parameter regimes of $\Delta$ and time intervals, confirming that our experimental arrays are sufficiently large to faithfully reproduce the semi-infinite dynamics.
To substantiate this claim, we consider the normalized cumulative deviation 
\begin{equation}
    C_N(z) = \frac{1}{z}\int\limits_0^zd\tilde{z}\,D_N(\tilde{z}).
\end{equation}
As stated above, we experimentally assess three sets of arrays, each corresponding to a specific value of $\Delta$.
Each set consists of ten arrays with varying values of the hopping distance $z_\mathrm{hop}$, made up of ten waveguides each (see Fig.~\ref{fig:setupsketch}).

The model Hamiltonian~\eqref{eq:ham} has units of length as it describes propagation along the $z$-axis of the WGA (see Fig.~\ref{fig:setupsketch}).
Consequently, we use $\beta z$ as dimensionless scale in the presentation of all subsequent results.
The observed evolution in these arrays is expected to cover time ranges of $\beta z < 4$. 
The numerical simulations depicted in Fig.~\ref{fig:deviation} predict values of $C_{10}(z=4/\beta)\sim10^{-6}-10^{-5}$ depending on the value of $\Delta$.
This means that the evolution of the truncated array differs only very slightly from a semi-infinite array within the considered parameter regime.

\textit{Measurement results}.\textemdash
We here present experimental data from three WGA sets, one for each of the three coupling regimes ($\Delta<1,\Delta\approx1,\Delta>1$).
Generally, we find good agreement between the experimental data and the single-band approximation discussed above.
However, with increasing coupling $\Delta$, deviations become more prominent.
We are able to attribute this to strong coupling effects, by showing that a numerical simulation of the full waveguide dynamics agrees consistently with the experimental data.

The first set (A1) corresponds to the regime exponential decay regime for $\Delta<1$, the second set (A2) to $\Delta=1$, and the third set (A3) to the bound-state regime $\Delta>1$.
Tab.~\ref{tab:measurementdata} collects their experimental parameters.
The numerical values of $\Delta$ in the different WGA sets were extracted from the experimental data by fitting simulated intensity distributions from a single-band model to the experimental data.
Following the approach of Ref.\ \cite{longhi2006nonexponential}, we introduce an effective decay rate to characterize deviations from exponential decay:
\begin{equation}\label{eq:decayrate}
    \gamma_\mathrm{eff}(t) = -\frac{\ln\left(|c_0(t)|^2\right)}{t}.
\end{equation}
In the case of purely exponential decay, where $c_0(t) = \exp(-\gamma_0 t/2)$, this expression yields $\gamma_\mathrm{eff}(t) = \gamma_0$. Therefore, any deviation of $\gamma_\mathrm{eff}(t)$ from a constant value signals a departure from exponential behavior.

\begin{figure}[h]
    \centering
    \includegraphics[width=\columnwidth]{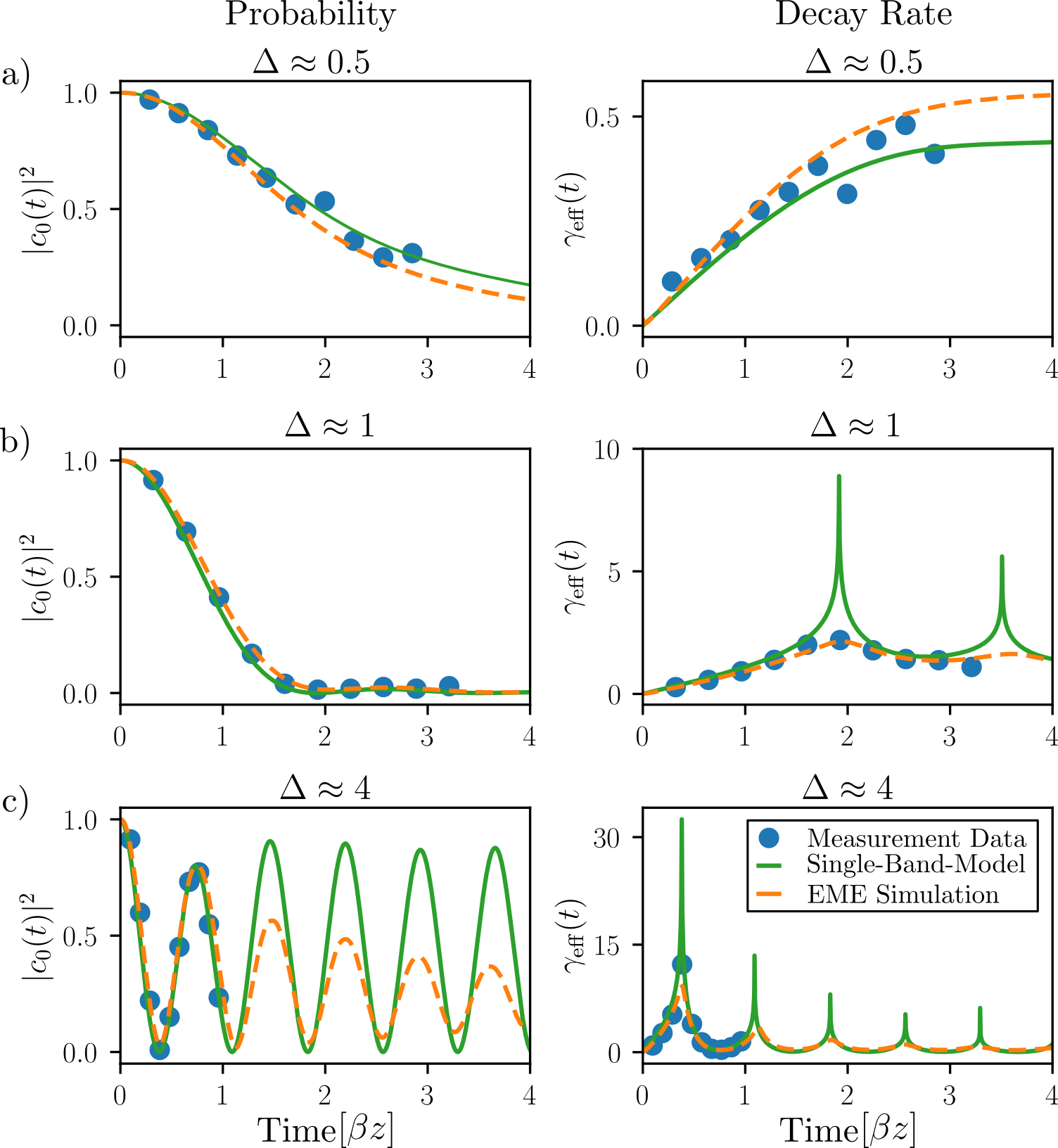}
    \caption{Experimental data (blue dots) for the population dynamics of the first waveguide, compared with the analytic model (green line) for the site occupation introduced in~\cite{longhi2006nonexponential} and the simulated population dynamics without a weak coupling assumption (orange dashed line).
    The left column shows the measured probabilities $|c_0(t)|^2$, and the right column the corresponding effective decay rates $\gamma_\mathrm{eff}(t)$ (see Eq.~\eqref{eq:decayrate}).
    Panel a) corresponds to $\Delta \approx 0.5$ (WGA set A1), b) to the uniform coupling case with $\Delta \approx 1$ (WGA set A2).
    Panel c) corresponds to case $\Delta > 1$, explicitly $\Delta \approx 4$ (WGA set A3).
    With a typical coupling strength of $\beta \approx 0.2 /\mathrm{cm}$ the time range of $4$ corresponds to a chip length of 20 cm.}
    \label{fig:evol}
\end{figure}

Fig.~\ref{fig:evol} presents the experimental data for the WGAs (A1)-(A3).
It compares the measured probabilities $|c_0(t)|^2$ and corresponding excitation rates  with the predictions from the single-band approximation of Eq.~\eqref{eq:c0}, and with exact numerical simulations of the WGAs.
The comparison demonstrates that for strong coupling the single-band approximation is not applicable, as it assumes orthogonal eigenmodes for each waveguide and only next-neighbor coupling.
Instead, when bringing the waveguides into close proximity for strong coupling, the system should be treated as one large potential with multiple supermodes.
To achieve this, we make use of the eigenmode expansion technique (EME).
As shown in the End Matter, this method is based on a refractive index reconstruction, so as to resemble each experimental probe as closely as possible.

In the case of $\Delta \approx 0.50$ (A1), we find that the experimental data qualitatively reproduces the theoretical predictions well.
The decay in the first waveguide closely follows the predicted behavior across all observed time steps, as shown in Fig.~\ref{fig:evol}a).
In this regime, $\gamma_\mathrm{eff}(t)$ exhibits convergence toward a constant value, indicating that the decay approaches an exponential form over time.
For short timescales, however, strong non-exponential behavior can be observed.
The EME simulation indicates that supermodes do not play a major role in the overall evolution.
The deviations between single-band approximation and EME at long evolution times are thus not attributable to supermodes, but rather to slight differences in the estimated coupling behavior arising from the two fundamentally different approaches.

\begin{table}[b!]
\centering
\begin{tblr}{
  vline{2} = {-}{},
  hline{3} = {-}{},
}
 & $d_0$ & $d$ & $\beta_0$ & $\beta$ & $\Delta$ & $\sigma_\mathrm{SB}$ & $\sigma_\mathrm{EME}$\\
 unit & $\mu \mathrm{m}$ & $\mu \mathrm{m}$ & $1/\mathrm{cm}$ & $1/\mathrm{cm}$ & $1$ & $1$ & $1$ \\ 
 A1 & $31.2$ & $27.1$ & $0.090$ & $0.190$ & $0.474$ & $0.035$ & $0.051$ \\
 A2 & $27.1$ & $27.1$ & $0.214$ & $0.214$ & $1.0$ & $0.025$ & $0.014$  \\
 A3 & $19.6$ & $27.1$ & $0.800$ & $0.192$ & $4.17$ & $0.050$ & $0.064$ 
\end{tblr}
\caption{Waveguide properties of the WGA sets. $d_0$ and $d$ are measured values of the observed mode positions. $\beta_0$ and $\beta$ and thus also $\Delta$ are the values which yield the best result when fitting coupled mode theory to the measured evolution. $\sigma_\mathrm{WC}$ and $\sigma_\mathrm{EM}$ are the root mean square errors between the measurement points and the analytical model and simulation of the first site respectively.}
\label{tab:measurementdata}
\end{table}

For the case of uniform coupling, \ie $\Delta \approx 1$ (A2), deviations in the decay rate from the idealized theoretical model become more pronounced.
As shown in Fig.~\ref{fig:evol}b), the decay is clearly non-exponential throughout the measured time interval.
The experimental behavior of $|c_0(t)|^2$ qualitatively matches the model's predictions.
The predicted peak in $\gamma_\mathrm{eff}(t)$, however, appears much less pronounced in the experimental data and in the results of the EME simulation.
This shows that the waveguide system deviates from the analytical model due to dynamics beyond the single-band model.

In the case of $\Delta > 1$ (A3), as shown in Fig.~\ref{fig:evol}c), both $|c_0(t)|^2$ and $\gamma_\mathrm{eff}(t)$ exhibit excellent agreement with the theoretical predictions except for the same shortcomings as described for the uniform coupling in (A2).
The strong coupling between the first and second sites leads to pronounced oscillations, resulting in distinctly non-exponential behavior.
In this strong-coupling case the largest deviations between the single-band model and the EME simulation can be seen especially for values of $\beta z > 1$, where the maxima are not as pronounced in the EME simulation.
Moreover, the minima are farther from zero as can be seen at $\beta z \approx 3.29$. 
These two effects reduce the contrast of the EME results, in comparison with the prediction from the single-band model.
This also causes the decay rate to have much less pronounced peaks in case of the EME simulation and measurements, even for the first minima of the probability at $\beta z \approx 0.38$.

As shown in Table \ref{tab:measurementdata}, the root mean square errors between theory and measurement, as well as between simulation and measurement, are  below $6.4\%$ in all cases.
Nevertheless, the error fluctuations, being the smallest for the uniform coupling, most likely stem from the manufacturing process. 
By considering all sites, instead of only the first, in the error calculation, we find that the root mean square errors between experiment and the single-band model amounts to
$\sigma_\mathrm{SB}^\mathrm{A1}=0.022$, $\sigma_\mathrm{SB}^\mathrm{A2}=0.027$ and $\sigma_\mathrm{SB}^\mathrm{A4}=0.021$.
The deviations between experiment and EME simulation are
$\sigma_\mathrm{EME}^\mathrm{A1}=0.035$, $\sigma_\mathrm{EME}^\mathrm{A2}=0.034$ and $\sigma_\mathrm{EME}^\mathrm{A3}=0.031$.
This indicates that the error between the two theoretical models and the experimental data remains nearly constant within our considered parameter range.
The single-band model agrees slightly better with the experimental results, because its two free parameters ($\beta,\Delta$) are fitted to the measured time evolution.
In contrast, the EME simulation is derived only from a fit to the single-mode profile (see End Matter).

\textit{Conclusions}.\textemdash
In summary, we have experimentally implemented and verified the dynamics of a prototypical boundary-defect model using photonic waveguide arrays. 
Over the considered propagation distances, these arrays are shown to faithfully reproduce the behavior of systems with effectively infinite degrees of freedom despite their finite size, including non-exponential decay and coherent oscillations.
By comparing the observed dynamics with exact analytical predictions, we demonstrated excellent quantitative agreement across all relevant coupling regimes. 
This confirms the reliability and flexibility of photonic waveguide arrays as a platform for simulating tight-binding models and for exploring dynamical features such as memory effects and open-system behavior. 
Our results highlight the potential of this approach for advancing analog quantum simulation and investigating foundational aspects of quantum transport in engineered systems.

\section{Acknowledgements}

We thank Harald Weinfurter for his unwavering support. 
His backing has been invaluable to the success of this work.
FHH, BB, CV and JM acknowledge funding by the German Research Foundation (Deutsche Forschungsgemeinschaft, DFG) under Germany’s Excellence Strategy—EXC-2111—390814868 and by the Bavarian Ministry for Science and the Arts under the project EQAP (CR 20211118). 
JK gratefully acknowledges support from Dr.~Max R\"{o}ssler, the Walter Haefner Foundation, and the ETH Z\"{u}rich Foundation.
RHJ gratefully acknowledges support by the Wenner-Gren Foundations and by the Wallenberg Initiative on Networks and Quantum Information (WINQ).
Nordita is supported in part by NordForsk.
AS acknowledges Funding from the Deutsche Forschungsgemeinschaft (grants SZ 276/9-2, SZ 276/19-1, SZ 276/20-1, SZ 276/21-1, SZ 276/27-1, and GRK 2676/1-2023 ‘Imaging of Quantum Systems’, project no.
437567992).
AS also acknowledges funding from the Krupp von Bohlen and Halbach Foundation
as well as from the FET Open Grant EPIQUS (grant no. 899368) within the framework of the
European H2020 programme for Excellent Science and funding from the
Deutsche Forschungsgemeinschaft via SFB 1477 ‘Light–Matter Interactions at Interfaces’ (project
no. 441234705).\\

FHH and BB contributed equally to this work.

\bibliography{Chained-Waveguides}

\section{End Matter} \label{endMatter}
\textit{Closed-form solution of $c_0(t)$}.\textemdash
As apparent from Eq.~\eqref{eq:c0}, the time evolution for the cases $\Delta<1$ and $\Delta>1$ is not only governed by an exponential decay and a sinusoidal oscillation, respectively.
In particular, $c_0(t)$ is modified by the terms
\begin{multline}
    S_{<}(t) = 2J_0(2t)+\left(1+\frac{1}{\gamma^2}\right)\cdot\\\cdot\Biggl(\frac{1}{2}\sum\limits_{l=-\infty}^{\infty}\frac{J_l(2t)}{\gamma^l}-\sum\limits_{l=0}^{\infty}\frac{J_{2l}(2t)}{\gamma^{2l}}\Biggl),
    \label{eq:S<}
\end{multline}
and 
\begin{multline}
    S_>(t) = J_0(2t)-(1-\frac{1}{\gamma^2})\cdot\\\cdot\Biggl(\sum\limits_{n = 0}^\infty\sum\limits_{l=n}^{2n}\frac{(it)^{2n}}{l!(2n-l)!}\gamma^{2l-2n}\Biggl),
    \label{eq:S>}
\end{multline}
where $J_l(x)$ are Bessel functions of the first kind, and $\gamma=\sqrt{|1-\Delta^2|}$. 
Note that in Eq.~\eqref{eq:S<}, all terms decay over time. 
Furthermore, apart from the term $\propto J_0(2t)$, all terms are suppressed by a factor of $1/\gamma$. 
Eq.~\eqref{eq:S>} shows similar behavior in this regard. 
However, we observe that this expression oscillates with large amplitudes for $\Delta\approx1$. 
In these cases, the term acts as a counterweight to the sinusoidal term of Eq.~\eqref{eq:c0} and removes most of the oscillatory behavior in favor of the remaining Bessel function which dominates the evolution in this regime. 
For $\Delta\gg1$, the roles are reversed: $S_>(t)$ is highly suppressed and the evolution is entirely governed by the sinusoidal term of Eq.~\eqref{eq:c0}.\\

\begin{figure}
    \centering
    \includegraphics[width=\columnwidth]{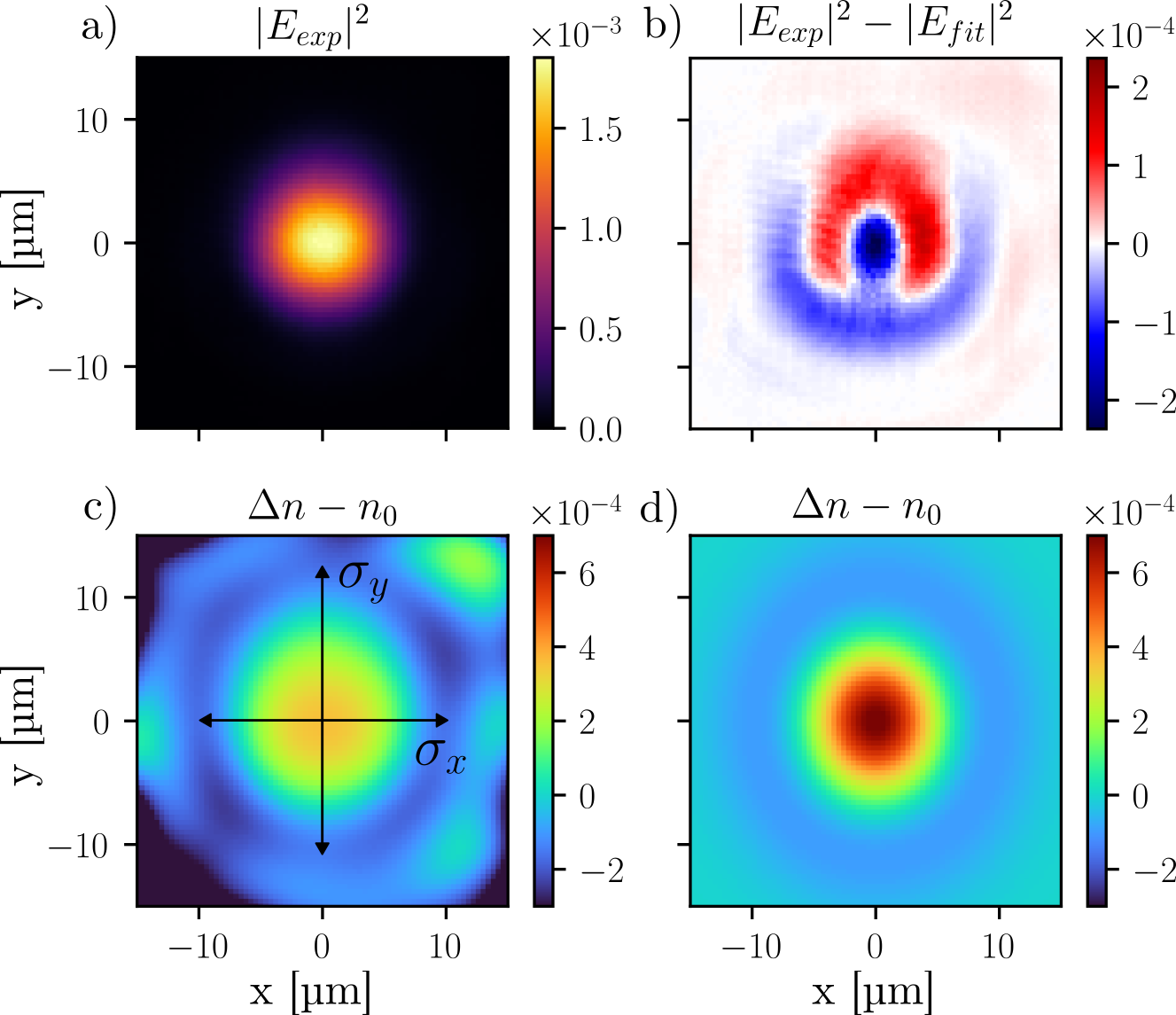}
    \caption{Reconstruction process of the refractive index profile of the waveguides for the EME simulation.
    All four images are transversal profiles in the x-y plane. Via the paraxial Helmholtz equation one can calculate from a measured (and normalized) intensity profile of an eigenmode a) back to the refractive index profile c) where $\Delta n$ with respect to the refractive index of the substrate is shown. With $\sigma_x$ and $\sigma_y$ from c) one can find the best fitting Ricker wavelet d), representing the analytical refractive index profile, generating an eigenmode closest to the measured image a). The deviations from this theoretical eigenmode to the measured image are shown in b) where the deviations are $<0.024\%$. The overall Fidelity between fit and measurement is $98.9\%$. The reconstructed refractive index is in good agreement with the observed evolution in the WGA.}
    \label{fig:helmholtz}
\end{figure}

\textit{Intensity measurement}.\textemdash
To evaluate our experimental data, based on the weak coupling approach, we use an approximate basis decomposition into localized modes. 
These modes are eigenmodes $\Phi(x,y)$ of an individual waveguide of the array, which have nearly no overlap between each other if the waveguides are well separated.
In comparison to this approximation however, the actual eigenmodes of the structure, are spread out over all waveguides of the array.
With this approximate basis decomposition, the intensity within each waveguide is given by the overlap of the measured electric field distribution $\Psi(x,y)$ and a single localized mode profile  $\Phi(x,y)$ centered at the waveguide's x-coordinate $d_i$,
\begin{equation}
    |c_i|^2 = \int dxdy\,| \Psi(x,y) \Phi(x-d_i,y) |^2. 
\end{equation}
For the profile $\Phi(x,y)$ we use the eigenmode generated by the refractive index profile of Fig.~\ref{fig:helmholtz}d).

Thereby, we assume a flat wavefront for all our measured intensities which allows the direct conversion between the electric field and intensity, $E_0\cdot\Psi(x,y) =E(x,y)= \sqrt{I(x,y)}$. 
\\

\textit{Refractive index reconstruction}.\textemdash
To conduct an accurate EME simulation the refractive index profile of the WGA is needed.
We reconstruct the total profile based on the profile of a single isolated waveguide.
For an isolated waveguide which carries only a single eigenmode the refractive index can be obtained from its electric field distribution based on the paraxial Helmholtz equation~\cite{alberucci2020indepth}
\begin{equation}
    k^2 \Psi(x,y) = \nabla^2 \Psi(x,y)+k_0^2 n(x,y)^2,
    \label{eq:paraxial}
\end{equation}
where $k_0$ is the free space wavenumber and $k$ is the wavenumber for the mode $\Psi(x,y)$ given by $\frac{2 \pi}{\lambda}n_\mathrm{eff}$ with $\lambda$ as the vacuum wavelength and $n_\mathrm{eff}$ as the effective, average refractive index of the eigenmode.
Reconstructing the refractive index $n(x,y)$ via a given electric field $\Psi(x,y)$ with this equation is extremely sensitive to noise and, therefore, is error-prone in areas with low intensity. 
Nevertheless, the reconstruction can still be used to gain information about the overall shape of the refractive index profile in areas with high intensity and allows for finding an analytical expression. 
As also shown in previous publications \cite{alberucci2020indepth, michele2019nearir}, the profile, as shown in Fig. \ref{fig:helmholtz}c), is similar to a 2D Ricker-Wavelet
\begin{equation}
    n(x,y) = \Delta n \left(1-\frac{2x^2}{\sigma_x^2}-\frac{2y^2}{\sigma_y^2} \right)e^{-\frac{2x^2}{\sigma_x^2}-\frac{2x^2}{\sigma_x^2}} + n_0,
\end{equation}
with $\Delta n$ as the peak amplitude of the refractive index change compared to the substrate $n_0$ and $\sigma_i$ as the width of the function in the respective direction $x$ or $y$.

A fit of the Ricker-Wavelet to the reconstructed refractive index however does not take the negative dip in the refractive index into account.
Therefore, we use the Helmholtz reconstruction only to get the values of $\sigma_x$ and $\sigma_y$ which is the distance between the minima in $x$ and $y$ direction to the center maximum.

$\Delta n$ is now fitted by finding the best overlap between the measured eigenmode and the reconstructed eigenmode that belongs to the refractive index profile of a Ricker-Wavelet with a fixed $\sigma_x$ and $\sigma_y$ and a varying $\Delta n$.
The eigenmode that belongs to a given refractive index profile can be found by treating Eq.~\eqref{eq:paraxial} as an eigenvalue equation and solving it numerically.
With this method we reliably reconstruct a refractive index profile that generates the measured eigenmode of a single waveguide as depicted in Fig.~\ref{fig:helmholtz}.\\

\textit{Eigenmode expansion (EME)}.\textemdash
For the EME simulation of the whole WGA system, we use the refractive index profile
$
    n_\mathrm{tot}=\sum_{i=1}^{10} n(x-d_i,y),
$
\ie  copies of the isolated profile determined above, with spacings as given in Table~\ref{tab:measurementdata}.
This profile has ten non-dissipative eigenmodes which define a basis for our waveguide system. 
The Gaussian input beam, centered on the first waveguide, is decomposed into the generated basis and propagated forward in distance (equivalently, in time).
The corresponding eigenvalues define the effective refractive indices, and thus the propagation velocities, of the respective modes.
This eigenmode-expansion method provides a realistic simulation of our experimental platform imitating it as close as possible as seen in Fig.~\ref{fig:evol}.

\end{document}